\documentstyle[twoside,fleqn,espcrc2,epsf,eqnfix]{article}
\def\h{\hfill\break}
\newcommand{\be} {\begin{equation}}
\newcommand{\ee} {\end{equation}}
\newcommand{\bdm} {\begin{displaymath}}
\newcommand{\edm} {\end{displaymath}}
\newcommand{\bc} {\begin{center}}
\newcommand{\ec} {\end{center}}
\newcommand{\beqa} {\begin{eqnarray}}
\newcommand{\eeqa} {\end{eqnarray}}

\newcommand{\bear}{\begin{eqnarray}}
\newcommand{\ear}{\end{eqnarray}}
\newcommand{\bea}{\begin{eqnarray*}}
\newcommand{\ea}{\end{eqnarray*}}

\def\ct{\cite}
\def\pmb#1{\setbox0=\hbox{#1}% \kern-.025em\copy0\kern-\wd0
\kern.05em\copy0\kern-\wd0 \kern-.025em\raise.0433em\box0 }

\def\ell{l}

\def\P{I\!\!P}

\def\P{I\!\!P}
\def\pd{\partial}

\def\indexname{Index}

\def\refname{References}
\def\b{\hskip -1truemm{\BrickRed{$\bullet ~$}}}
\input colordvi

\font\fiverm=cmr5
\font\sevenrm=cmr7

\newcommand{\AmS}{{\protect\the\textfont2
   A\kern-.1667em\lower.5ex\hbox{M}\kern-.125emS}}

\title{Soft and hard QCD
\thanks{This talk is taken largely from my new book with Donnachie, Dosch and
Nachtmann\cite{book}}}
\author{P V Landshoff%
\address{Centre for Mathematical Sciences \\
         Cambridge CB2 0WA \\
         pvl@damtp.cam.ac.uk}%
}

\begin{document}

\begin{abstract}\noindent
I review various theoretical questions that arise from data from
HERA and the Tevatron, and which are relevant for the LHC. They
range from soft physics, such as the total cross section, to
hard physics, such as Higgs production. In particular, I argue
that the proton's gluon density is somewhat larger at small $x$
than is currently accepted.
\end{abstract}

% typeset front matter (including abstract)
\maketitle

\begin{figure}[p]
\bc
{\epsfxsize=0.9\hsize\epsfbox[90 580 330 775]{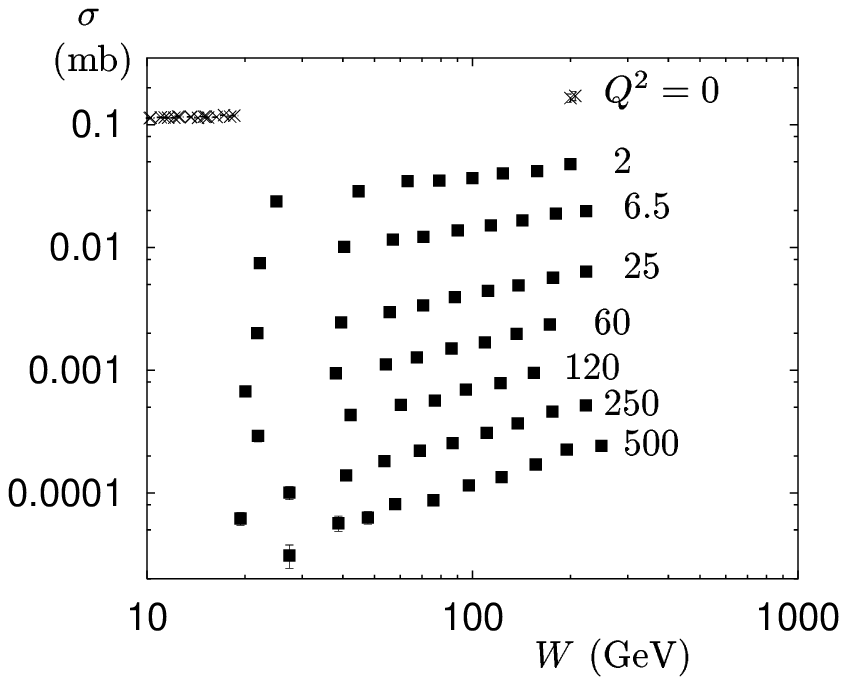}}
\ec
\vskip -9truemm
\caption{Data\ct{H101} for $\sigma^{\gamma^* p}$
at various values of $Q^2$,
together with the real-photon data of figure~\ref{GAMMAP}}
\label{DISSIG}
\vskip 9truemm
\bc
\epsfxsize=\hsize\epsfbox[60 580 320 760]{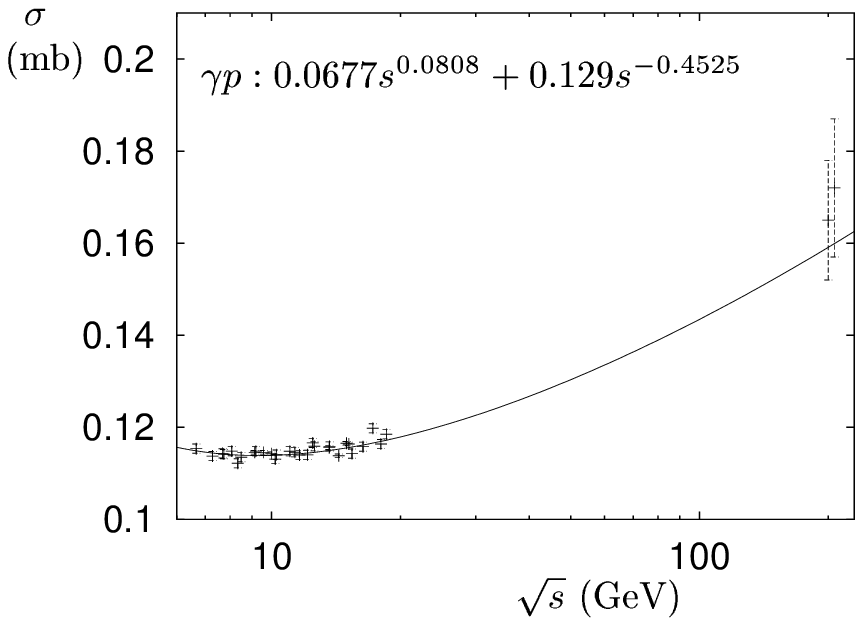}
\ec
\vskip -9truemm
\caption{$\gamma p$ total cross section; the curve\cite{DL92} takes account of %
the exchange of the soft pomeron, $f_2$ and $a_2$}
\label{GAMMAP}
\vskip 9truemm
\bc
\epsfxsize=0.8\hsize\epsfbox[0 0 330 115]{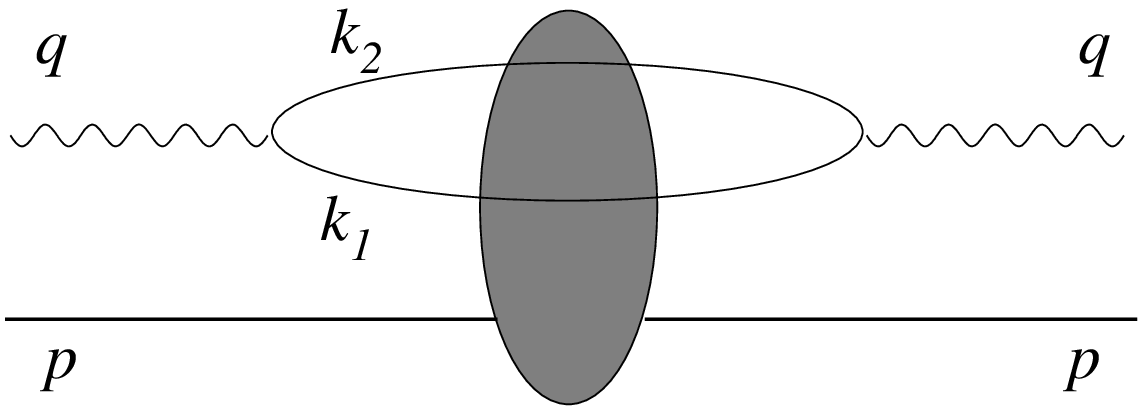}
\end{center}
\vskip -9truemm
\caption{Dipole model: interaction of a proton  with a highly-energetic photon}
\label{FDIS}
\end{figure}

\section{Introduction}
There are many approaches to high-energy scattering, among them

\b Regge theory

\b dipole models

\b stochastic vacuum models

\b saturation models

\b semiclassical approach

\b effective field theory

\b DGLAP

\b BFKL

They all use different language, but there are many links between them.
At present, none of them offers an agreed fundamental explanation for
the very striking discovery at HERA, that at high $Q^2$ the $\gamma^*p$
total cross section rises dramatically with increasing energy $W$.
This is seen in figure~\ref{DISSIG}. At small $Q^2$ the rise is
compatible with that seen in hadron-hadron total cross sections,
$(W^2)^{\epsilon_1}$ with $\epsilon_1\approx 0.08$, but at high
$Q^2$ the effective power is close to 0.4. Leading-order BFKL predicted this,
but unfortunately there is\cite{FL98,CC97} a huge correction in
next-to-leading order in $\alpha_s$. 

\section{Difficulty with DGLAP}

\def\u{{\bf u}}\def\P{{\bf P}}\def\f{{\bf f}}
Most fits to the data achieve the
rising power from DGLAP evolution\cite{H101,zeusfit,mrst,cteq}, but they do
so by making an expansion of the splitting matrix that is 
mathematically illegal.

The singlet DGLAP equation introduces a two-component quantity
\be
\u(x,t)= \left (\matrix{x\sum _f(q_f+\bar q_f)\cr xg(x,t)\cr}\right )
~~~~t=\log (Q^2/\Lambda^2)
\ee
If one Mellin transforms with respect to $x$
$$
\u(N,Q^2)=\int _0^1dx\,x^{N-1}\u(x,Q^2) ~~~~~~~~~~~~~~~~~~~~~~~~~~~~~~~~~~~~~~~~~~~~~~~~~~~~~~~~~~~~~~~~~~~
$$
\be
\P (N,\alpha_s(Q^2))=\int_0^1 dz\,z^N\P (z,\alpha_s(Q^2))
\ee
the equation becomes very simple:
\be
\BrickRed{{\pd\over\pd t}\u (N,Q^2)= \P (N,\alpha_s(Q^2))\,\u(N,Q^2)}
\label{dglap}
\ee
The standard approach is to expand the splitting matrix $\P$ is powers
of $\alpha_s$, but this is \Blue{{invalid}} at small $N$.
Compare
\be
\sqrt{N^2+\alpha_s}-N=\alpha_s/2N-\alpha_s^2/8N^3+\dots
\ee
Here each term in the expansion is singular at $N=0$ but the function
itself is not: the expansion is illegal\cite{cudell} when $N$ is small.
I will discuss later how one might partially  overcome this difficulty.

\section{Dipole model}

Figure~\ref{FDIS} shows the virtual forward compton amplitude. Each virtal
photon couples to a quark-antiquark pair which may be regarded as a
colour dipole. This leads to\goodbreak 
$$
\sigma_{T,L}^{\gamma^*p}(x,Q^2)\,=~~~~~~~~~~~~~~~~~~~~~~~~~~~~~~~~~~~~~~~~~~~~~~~~~~~~~~~~~~~$$$$
\int d^2R\, dz\;\; \psi^*_{T,L}(Q,R,z)\; \sigma^{\hbox{\sevenrm dip}} (x,R)\;
\psi_{T,L}(Q,R,z) $$
\be
\ee
Here, $T,L$ denote the polarisation of the photon. $\psi_{T,L}(Q,R,z)$
is the wave function at the vertex that couples it to the $q\bar q$
pair; it depends on the transverse separation $R$ of the pair and
on the longitudinal momentum fraction $z$ of the quark.
$\sigma^{\hbox{\sevenrm dip}} (x,R)$ is the cross section for the
interaction of the colour dipole with the target proton.

\begin{figure}[p]
\bc
\epsfxsize=\hsize\epsfbox[160 605 420 770]{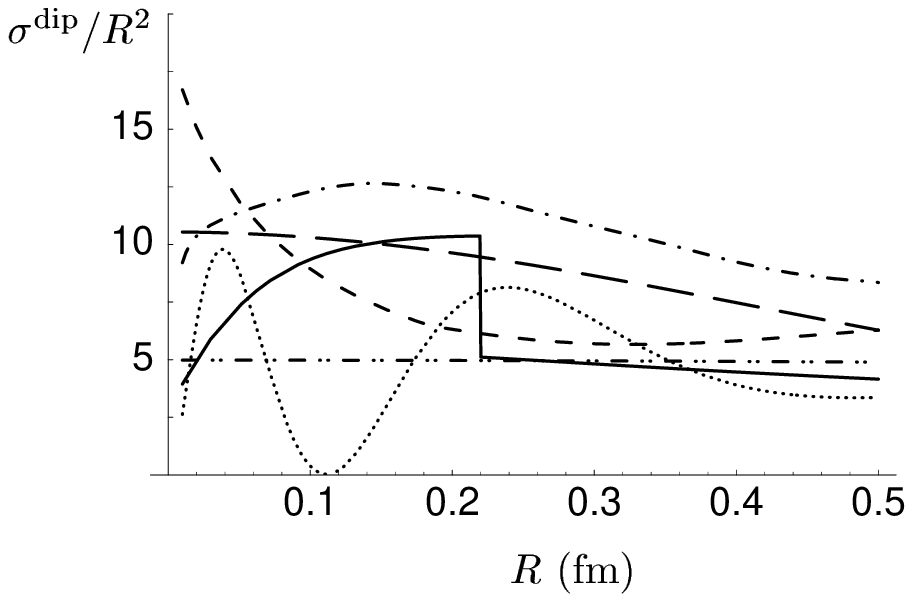}
\ec
\vskip -9truemm
\caption{Various forms for the dipole cross section\cite{book}}
\label{DIPCROSS}
\vskip 7truemm
\bc
\epsfxsize=0.7\hsize \epsfbox[0 0 450 290]{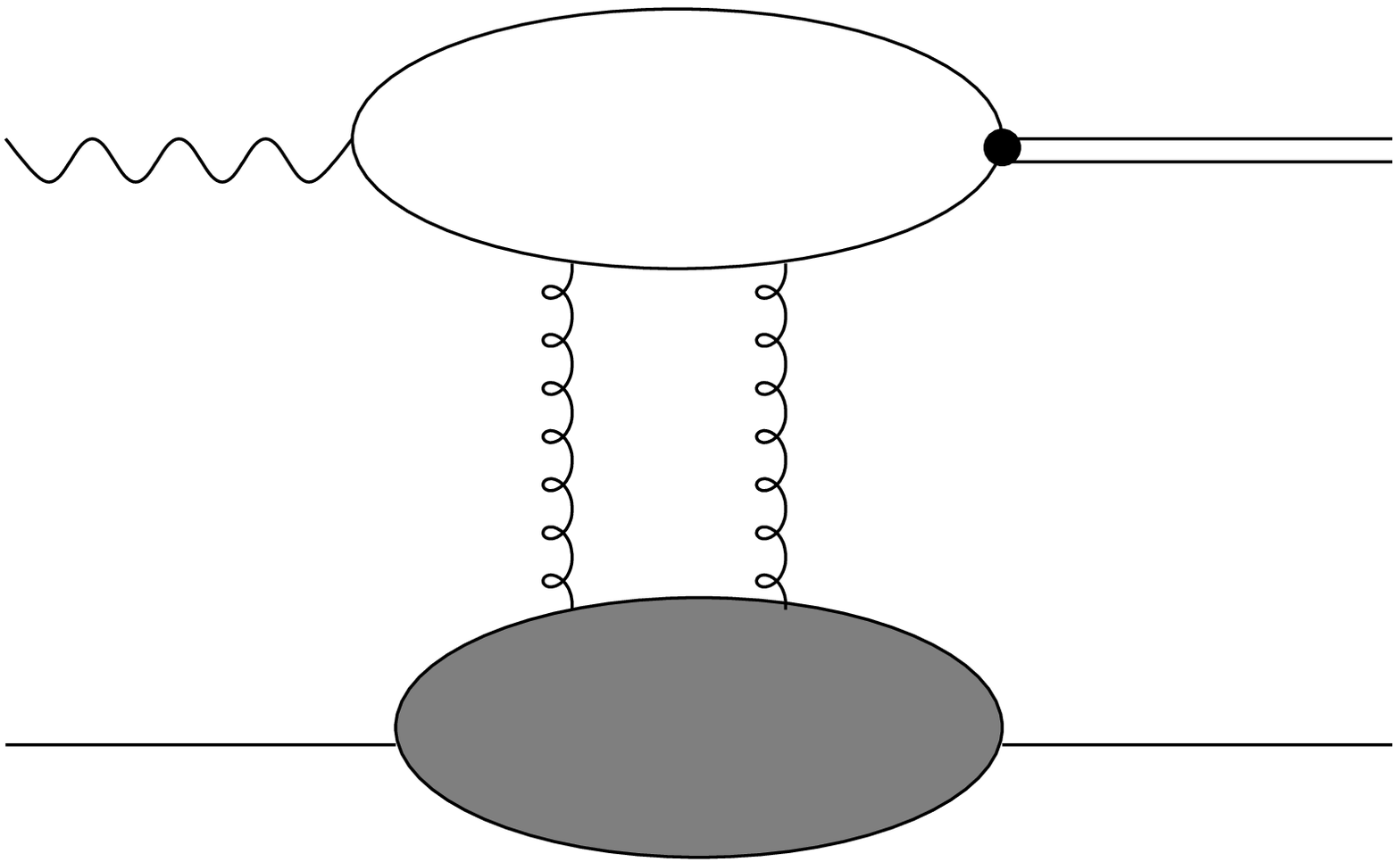}
\ec
\vskip -9truemm
\caption{Two-gluon exchange contribution to the $\gamma^*p\to Vp$
amplitude}
\label{VEC2GLUE}
\vskip 11truemm
\bc
{\epsfxsize \hsize\epsfbox[100 420 355 590]{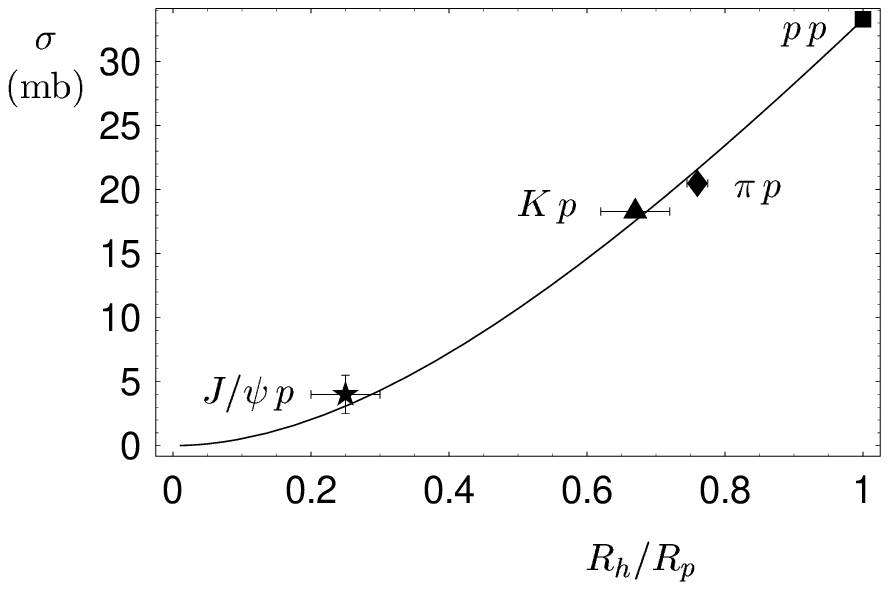}}
\ec
\vskip -9truemm
\caption{Relation between cross sections for various hadrons scattering
on a proton, and their radii}
\label{XSECHAD}
\end{figure}

The literature includes many different choices for 
$\sigma^{\hbox{\sevenrm dip}} (x,R)$. Figure~\ref{DIPCROSS} shows
a few of them. 
\RoyalPurple{There is no single dipole model!}

Nowadays, it is popular to combine the dipole model with the notion
of saturation. In hadron-hadron cross sections, such as $\sigma^{pp}$,
the Froissart-Martin-Lukaszuk bound\cite{Fro61,LM67} is familiar:
\be
\sigma^{pp}(s)\leq {\pi\over m^2_{\pi}}\log^2(s/s_0)
\ee
It is not a material constraint, because it gives an upper limit of
several \Blue{barns} at LHC energies! It is derived from unitarity:
\be
\hbox{Im }a_{\ell}(s)\geq |a_{\ell}(s)|^2
\ee
For $\gamma p$ scattering there is no similar inequality because
the unitarity relation does not contain an elastic term: to lowest order in
$\alpha_{\hbox{\sevenrm EM}}$ only hadronic final states are included.
In principle,
$\sigma^{\gamma p}(W)$ can get huge at large $W$ and
$F_2(x,Q^2)$ can get huge at small $x$
Nevertheless, many people believe that a Froissart-like bound is
saturated at an accessible $W$ or $x$,
and they implement this by writing an eikonal-like form
\be
\sigma^{\hbox{\sevenrm dip}} (x,R)=1\,-\,\exp\left(-{R^2}/{4 R_0^2(x)}\right )
\ee
Now
$$
1\,-\,\exp\left(-{R^2}/{4 R_0^2(x)}\right)=
{R^2\over 4 R_0^2(x)}-{R^4\over 32 R_0^4(x)}+\dots
$$
\be
\ee
and it is natural to identify the first term in this expansion
with the most elementary exchange, taken to be two-gluon exchange.
Figure~\ref{VEC2GLUE} shows this for the reaction $\gamma^*p\to Vp$. 
At the bottom of the figure there is
the proton's gluon structure function $xg(x,\mu^2)$, and so we have
for the dipole cross section 
$$
\sigma^{\hbox{\sevenrm dip}} (x,R)=~~~~~~~~~~~~~~~~~~~~~~~~~~~~~~~~~~~~~~~~~~~~~~~~~~~~~~~~~~~~~~~~~~~~~~~~~~~~~
$$
\be
\sigma_0\,\left\{
1\,-\,\exp\left(-{\pi^2\,R^2\,\alpha_s(\mu^2)\,xg(x,\mu^2)
\over 3\,\sigma_0}\right) \right\}
\ee
with just one parameter $\sigma_0$. Because $xg(x,\mu^2)$ obeys
DGLAP evolution, this model combines the dipole model with both
saturation and DGLAP, and it can give a good fit to experiment\cite{BGK02}.
However, there are many ways in which one can successfully fit the DIS
data.

\section{Stochastic vacuum model}

The stochastic vacuum model starts from the familiar vacuum gluon
condensate\cite{SVZ79} 
\be
\langle 0|:g^2F^{\mu\nu}(x)F_{\mu\nu}(y):|0\rangle\Big\arrowvert _{y=x} =M_c^4
\ee
with $M_c$ a few hundred MeV,
and generalises this relation to $y\not= x$.
This introduces a vacum correlation length. Some rather technical 
manipulations are needed, for example using the nonabelian Stokes
theorem\cite{DFK94}.
A particular realisation of the dipole model results, where 
the soft pomeron is generated from multigluon exchange.
The model successfully relates total cross sections to hadron sizes,
as is seen in figure~\ref{XSECHAD}

\section{Soft cross sections}

\begin{figure}[p]
\bc
\vskip -38truemm
\epsfxsize=0.9\hsize\epsfbox[70 590 310 760]{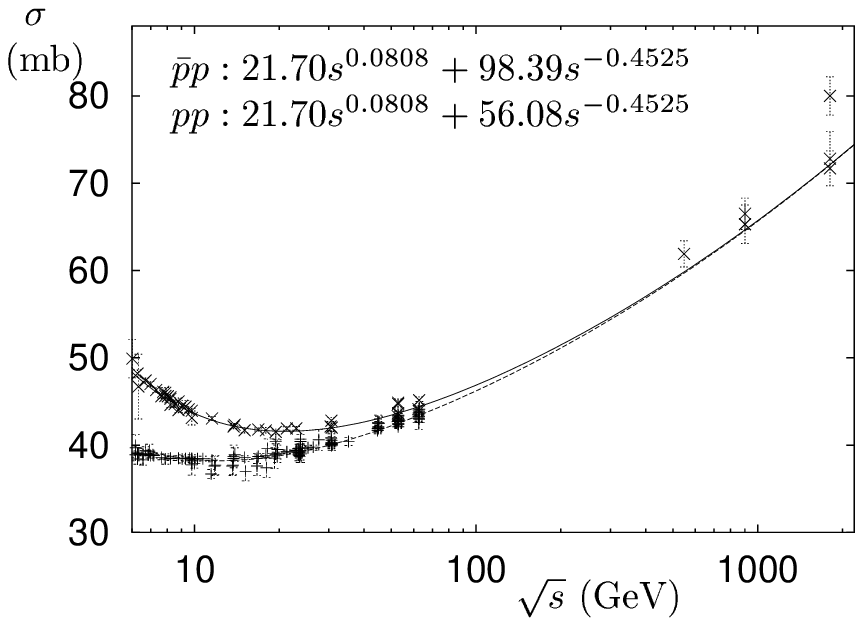}
\ec
\vskip -8truemm
\caption{$pp$ and $p\bar p$ total cross section data, with Regge fit\cite{DL92}}
\label{PP}
\vskip 7truemm
\bc
\epsfxsize=0.8\hsize\epsfbox[115 585 350 770]{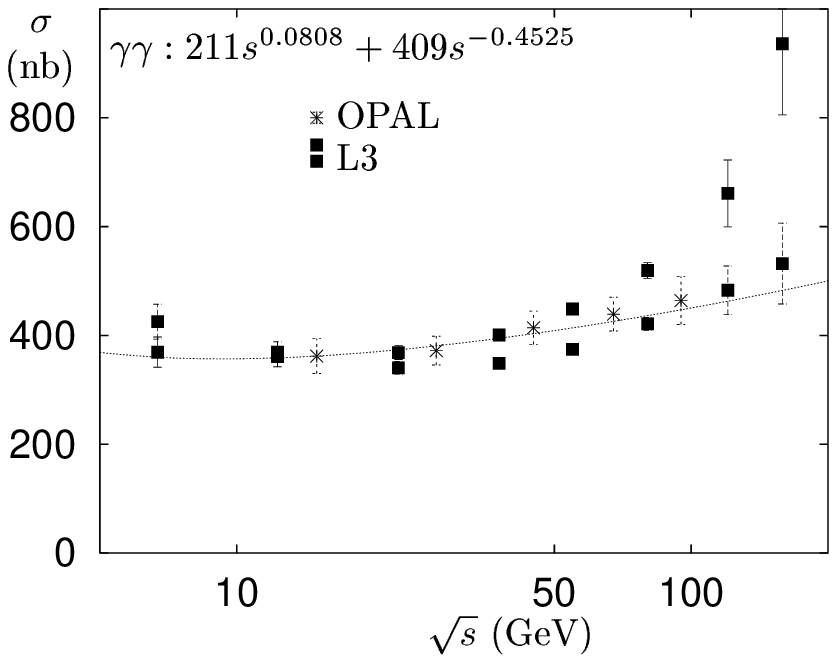}
\end{center}
\vskip -10truemm
\caption{$\gamma \gamma$ total cross section\ct{OPAL00,L301} with the
prediction obtained from factorisation}
\vskip -11truemm
\label{GG}
\vskip 19truemm
\bc
\epsfxsize=0.8\hsize\epsfbox[95 595 330 755]{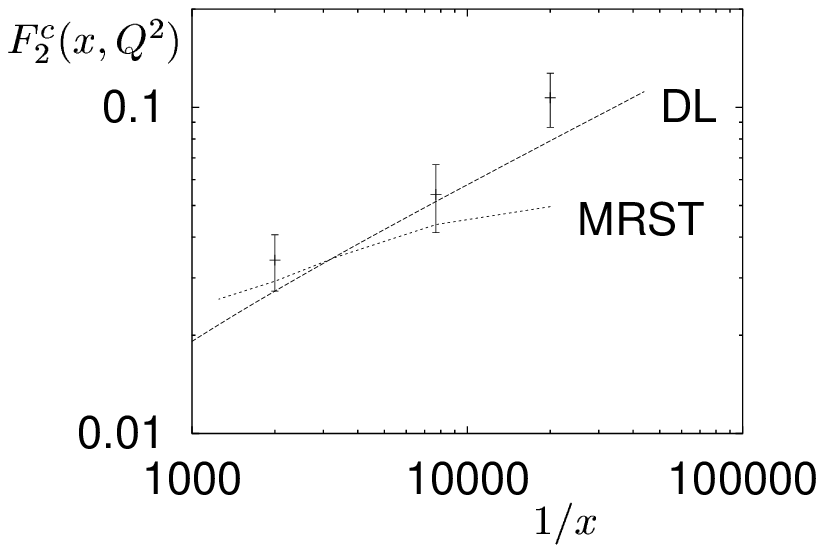}
\ec
\vskip -8truemm
\caption{Data\cite{Z00} for $F_2^c(x,Q^2)$ at $Q^2=1.8$ GeV$^2$. The curves
are the latest MRST fit\cite{mrst} and  a fit that includes only a 
hard-pomeron term\cite{DL01}}
\vskip -39truemm
\label{HARDCHARM}
\end{figure}

Regge theory provides remarkably simple fits to 
data for all hadron-hadron total cross sections\cite{DL92}. An
example is shown in figure~\ref{PP}. One needs just
two powers of $s$. One is close to $1/\sqrt s$ and is identified as
resulting from $\rho,\omega,f_2,a_2$ exchange. The other is close to
$s^{0.1}$ and its origin is unknown; to give it a name, 
we say that this term results from soft-pomeron exchange.
The extrapolation of the parametrisation shown in figure~\ref{PP} gives
108~mb at $\sqrt s=20$ TeV. Although, as I have explained, one can never hope
to achieve an energy at which the Froissart bound becomes a relevant
constraint, it has often led people to prefer to parametrise the rising
component of the cross section with a log$^2s$ term rather than a power. 
It is interesting that the most recent such fit\cite{compete} predicts
that the cross section at $\sqrt s=20$ TeV
will be some 10 mb greater than given by the simple power fit.

Note the highest-energy points in figure~\ref{PP}: 
the CDF point\cite{CDF94} is significantly
higher than the E710 point\cite{Amo89}. If CDF were to turn out to be
correct, this would signal the onset of some new term which
would significantly increase the cross section measured at the LHC.
The question whether such a hard term is present is of some fundamental
importance for the interpretation of the HERA data in figure~\ref{DISSIG}.
These data show clearly the presence of a hard term at high $Q^2$
and it is generally agreed that it should be understood through pQCD
evolution. But does the evolution \Blue{{\it{generate}}} the term, or merely
enhance its importance as $Q^2$ increases? I am fairly sure that it is
the latter that is the case\cite{cudell}. If this is true, the term should be
present in $\gamma p$ collisions already at $Q^2=0$. While there is
room for such an additional term in the data shown in figure~\ref{GAMMAP},
the error bars are too large to decide. The LEP data\cite{OPAL00,L301}
for the $\gamma\gamma$ total cross section are similarly
unclear. Figure~\ref{GG} shows the data. The L3 experiment presents two
sets of points, corresponding to two different Monte Carlos, which are
needed to correct for the fact that the detector's acceptance
is such that only a small fraction of the interactions are visible.
The curve represents a sum of the same two powers as the curves in
figure~\ref{PP}. The clearest indication that a hard term
is indeed present at small $Q^2$ is in the ZEUS\cite{Z00} data for the charm
structure function of the proton. As figure~\ref{HARDCHARM} shows,
already at $Q^2=1.8$ GeV$^2$ the rise with increasing $1/x$ of
$F_2^c(x,Q^2)$ is as rapid as that of the complete $F_2(x,Q^2)$ at large $Q^2$.
The same is even true at $Q^2=0$.
I return to this very important point later.

\begin{figure}[p]
\begin{center}
\vskip -6truemm
\epsfxsize=0.95\hsize\epsfbox[55 550 366 770]{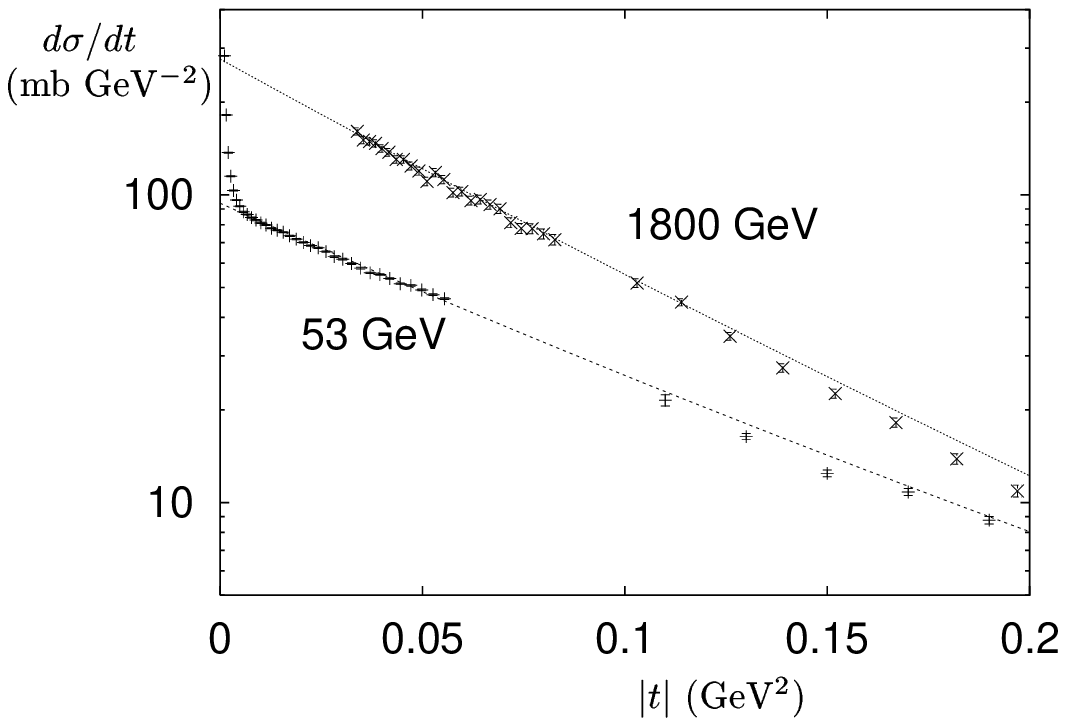}
\end{center}
\vskip -11truemm
\caption{$pp$ elastic scattering\cite{Nag79} at $\sqrt s=53$ GeV
and\ct{Amo90} $~\bar pp$ at $\sqrt{s}=1800$ GeV, together with
the curves (\ref{differential})
corresponding to $\alpha'_{\P}=0.25$ GeV$^{-2}$}
\label{SHRINKAGE}\index{elastic scattering!$pp$}
\vskip 4truemm
\begin{center}
\epsfxsize=0.95\hsize\epsfbox[66 604 300 770]{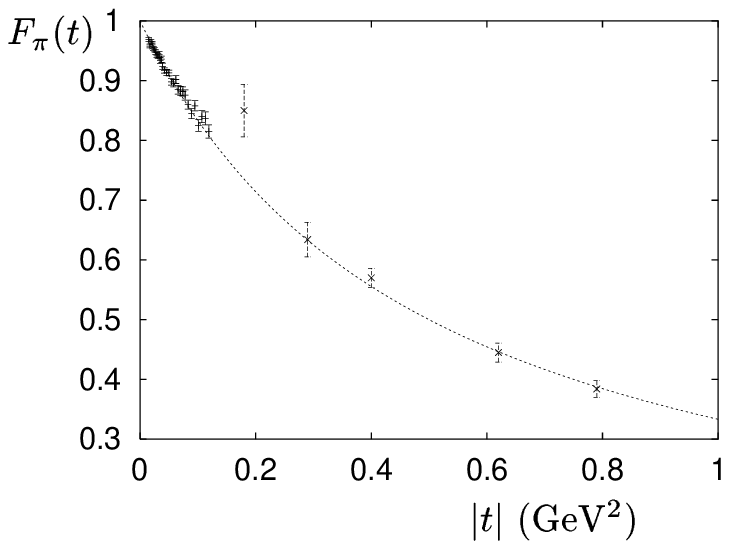}
\end{center}\index{form factor!pion}
\vskip -11truemm
\caption{Data\cite{Beb78,Ame84} for the pion elastic form factor, with the 
simple fit described in the text}
\label{PIONFF}
\begin{center}
\hskip -5truemm\epsfxsize=\hsize\epsfbox[66 604 300 770]{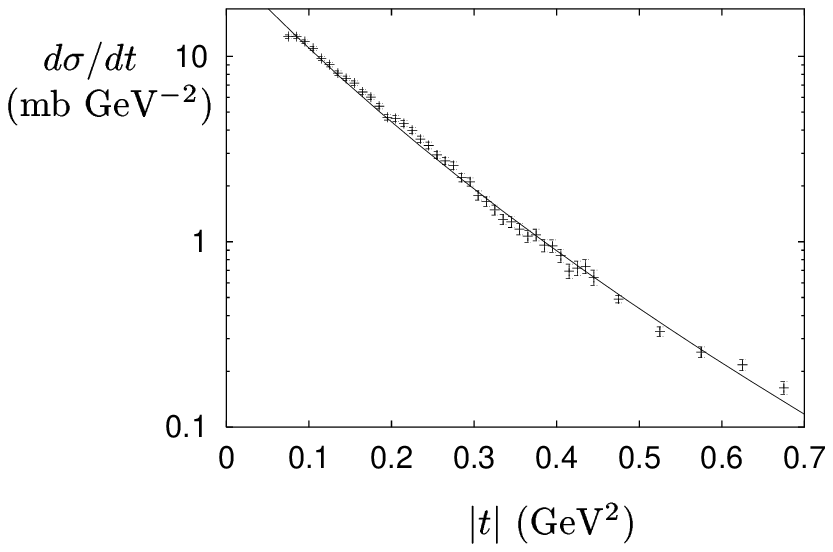}
\end{center}
\vskip -11truemm
\caption{$\pi p$ elastic scattering data\cite{Ake76} at \hbox{$\sqrt{s} = 19.4$ GeV}
with the curve (\ref{pidcs})}
\label{PIDCS}
\end{figure}
\begin{figure}[t]
\vskip -15truemm
\begin{center}
\epsfxsize=0.9\hsize\epsfbox[70 600 290 760]{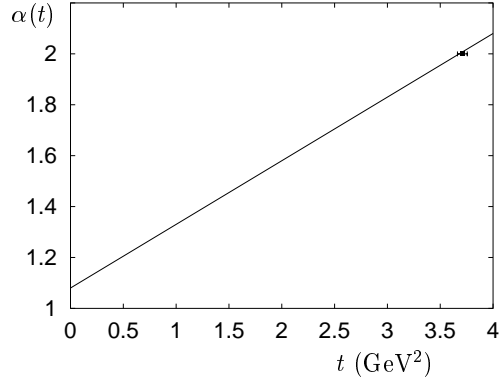}
\end{center}
\vskip -9truemm
\caption{A $2^{++}$ glueball candidate\ct{WA91}, with the line
$\alpha(t)=1.08+0.25t$}
\label{GLUEBALL}
\end{figure}

\section{Elastic scattering}

Regge theory provides a very simple extension  to elastic scattering 
of the total-cross-section
fit of figure~\ref{PP}. At high $s$, where only
the soft-pomeron term survives\cite{JL74,DL86},
\be
{d\sigma^{pp}\over dt}\sim {d\sigma^{\bar pp}\over dt}\sim {(3\beta_{\P}F_1(t))^4\over 4\pi}\left ({s\over s_0}\right )^{2\alpha_{\P}(t)-2}
\label{differential}
\ee
where
\be
\alpha_{\P}(t)=1.08+\alpha't
\label{alphapom}
\ee
and $F_1(t)$ is the proton's Dirac elastic form factor. The value
$\alpha'=0.25$ GeV$^{-2}$ is fixed by fitting to very accurate ISR
data at very small $t$. The form (\ref{differential})
then successfully predicts the data
at much higher energy. See figure~\ref{SHRINKAGE}.

With no free parameters, we may extend this to $\pi p$ elastic scattering.
The pion has only two valence quarks, so we replace (\ref{differential})
with 
\be
{d\sigma\over dt}={(2\beta_{\P}F_{\pi}(t))^2
(3\beta_{\P}F_1(t))^2\over 4\pi}\left ({s\over s_0}\right )
^{2\alpha_{\P}(t)-2}
\label{pidcs}
\ee
Figure~\ref{PIONFF} shows data for the pion form factor; they fit well to
$F_{\pi}(t)={1/(1-t/m_0^2)}$ with $m_0^2=0.5$ GeV$^2$.
This leads to the zero-parameter fit shown in figure~\ref{PIDCS}. 

Regge theory is remarkably successful!

\section{Glueballs}

\begin{figure}[p]
\bc
\vskip -1truemm
\epsfxsize=0.9\hsize\epsfbox[60 585 290 760]{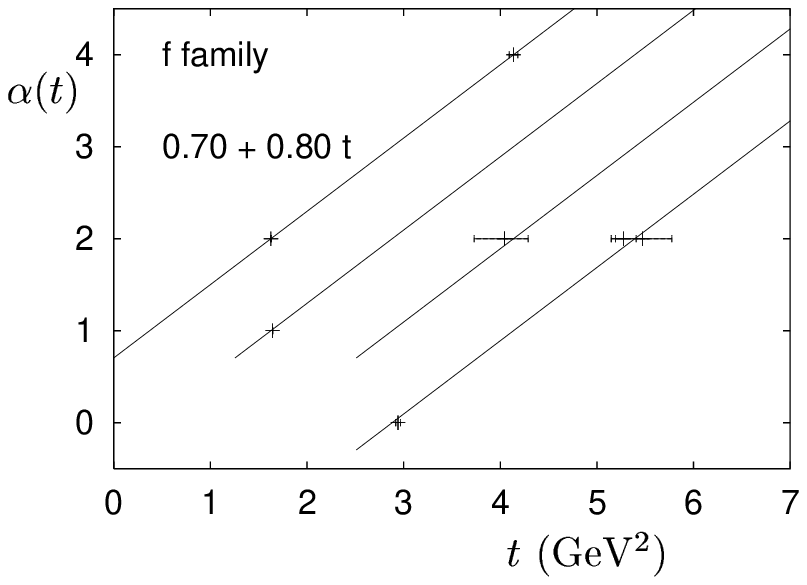}
\ec
\vskip -16truemm
\caption{Particles of the $f$ family. Only confirmed
states\cite{PDG00} are shown.}
\vskip -6truemm
\label{DAUGHTERS}
\vskip 9truemm
\begin{center}
\epsfxsize=1.05\hsize\epsfbox[35 575 320 775]{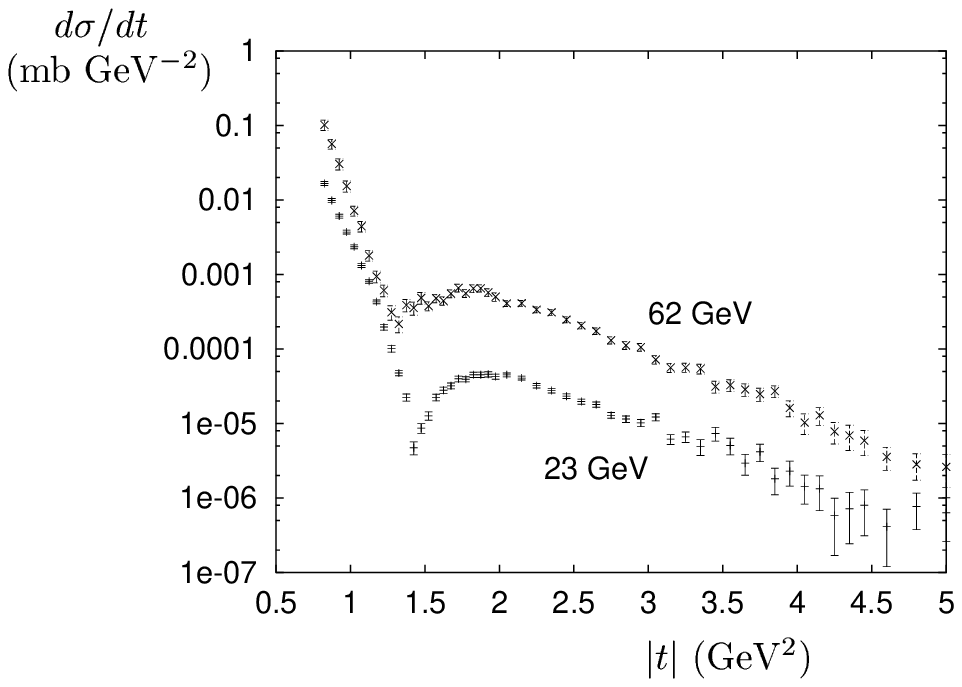}
\end{center}
\vspace{-13mm}
\caption{$pp$ elastic scattering data at large $t$
(CHHAV collaboration\cite{Nag79}). The 62 GeV data are multiplied by 10.}
\label{LARGEPP}
\vskip 4truemm
\begin{center}
\epsfxsize=0.9\hsize\epsfbox[50 580 310 770]{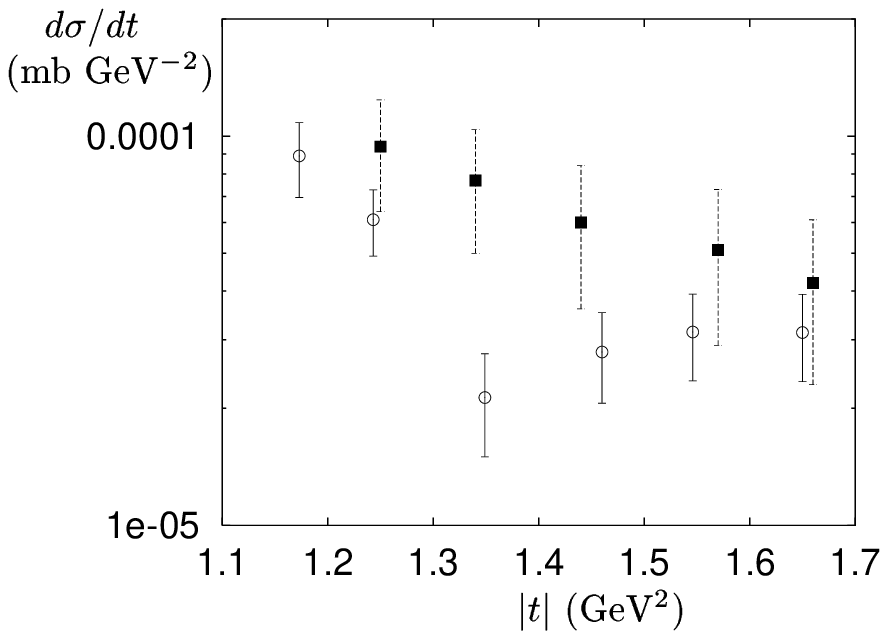}
\end{center}
\vskip -13truemm
\caption{Elastic scattering at $\sqrt{s}=53$~GeV of antiprotons
(upper points) and protons (lower points) on protons\ct{Bre85}}
\label{DIP}
\end{figure}

Although we do not understand the origin of the soft pomeron, there
is a wide feeling that it is just gluon exchange. If that is so,
and if its trajectory really is straight, as written in (\ref{alphapom}),
then the value of $t$ for which it passes through 2 should be the
square of the mass of a $2^{++}$ glueball. The WA91 collaboration\cite{WA91}
has a $2^{++}$ candidate of \Blue{exactly} the right mass: see figure~%
\ref{GLUEBALL}.

Nowadays it is known that Regge trajectories corresponding to ordinary
particles are accompanied by daughter trajectories\cite{Col77}.
These are trajectories separated by an integral number of units
from the parent trajectory. An example is the $f$ family, shown in
figure~\ref{DAUGHTERS}. The existence of daughters was predicted from
Regge theory at a time when little was known about the meson spectrum.
One would expect the pomeron trajectory to have daughters too. 
The search for glueballs is very important to give more understanding
about the pomerons -- the hard pomeron is probably associated with glueballs
too. 

\section{Odderon}

The minimum number of gluons  needed to model the pomerons is two,
because they represent colourless even-parity exchange.
With three gluons, one can model colourless odd-parity exchange,
called odderon exchange. There is a clear sign of odderon exchange
in $pp$ and $\bar pp$ elastic-scattering data at large $t$, but the mystery
is that so far odderon exchange has not been identified at $t=0$.

Figure~\ref{LARGEPP} shows ISR data for $pp$ elastic scattering. There is
a very striking dip at $|t|\approx 1.4$ GeV$^2$. The very last week of
running of the CERN ISR showed that $\bar pp$ elastic scattering is different:
the dip is filled in, as is seen in figure~\ref{DIP}.

\begin{figure}[t]
\begin{center}
\epsfxsize=\hsize\epsfbox[88 460 475 760]{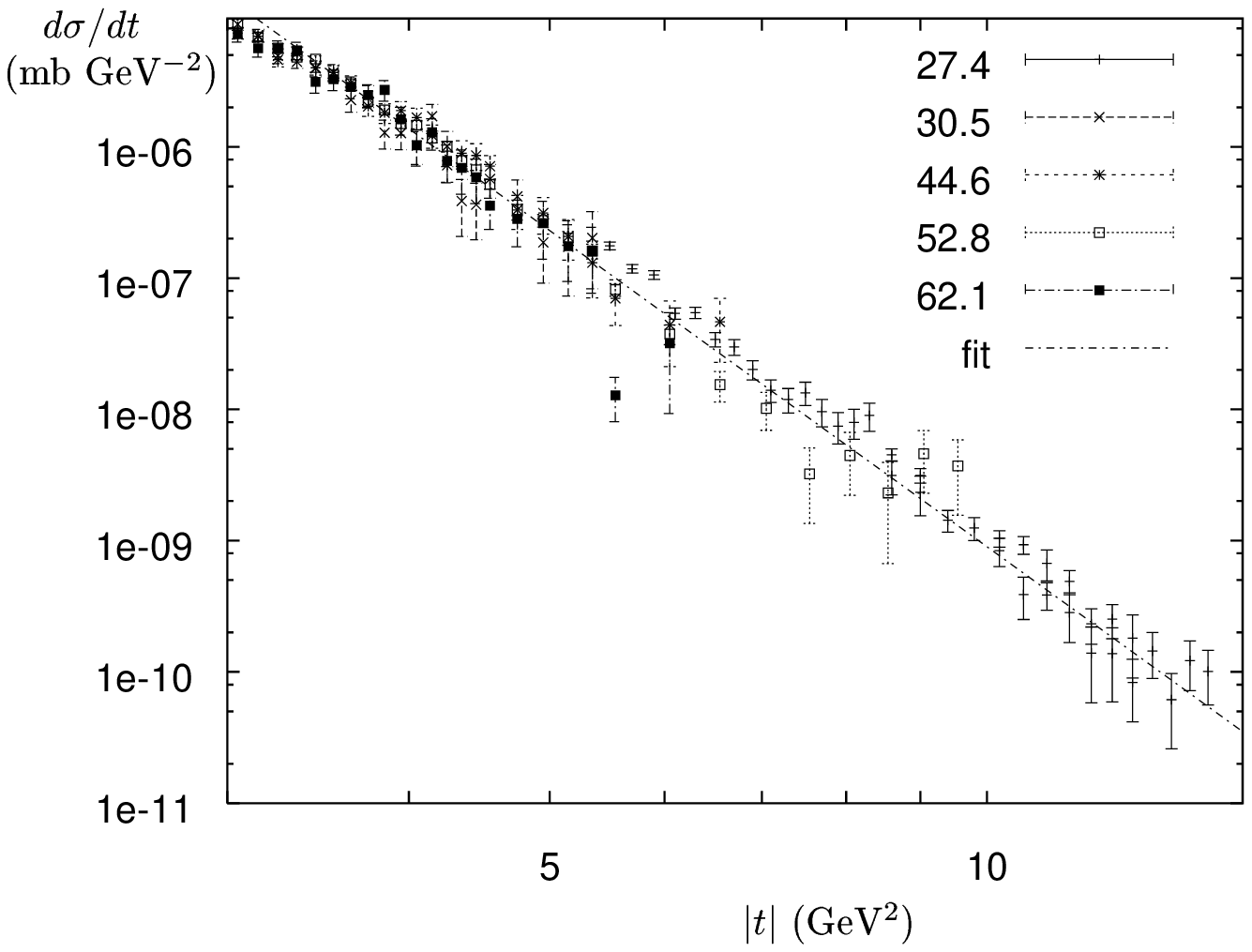}
\end{center}
\vskip -9truemm
\caption{$pp$ elastic scattering data\cite{Nag79,Fai81}
at the largest available $t$, at various energies indicated
in the figure as $\sqrt{s}$ in GeV. The line is $0.09\,t^{-8}$.}
\label{LARGE-T}
\vskip 9truemm
\bc
\epsfxsize=0.8\hsize\epsfbox[0 0 480 340]{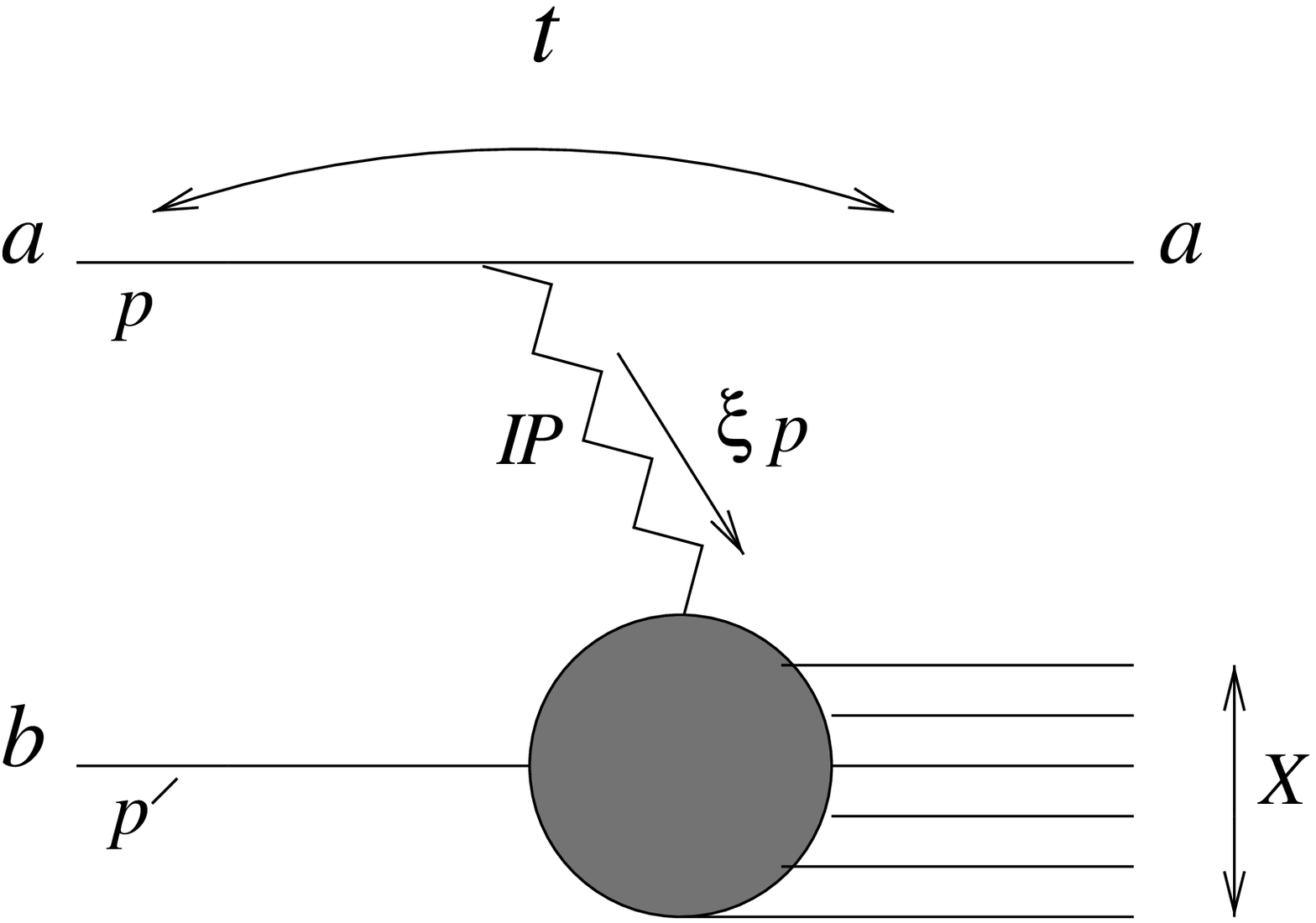}
\ec
\vskip -9truemm
\caption{Pomeron exchange in an inelastic diffractive event}
\label{DIFFRACTION}
\end{figure}

Beyond the dip, the data in the ISR energy range
fit very well to perturbative 3-gluon exchange
calculated\cite{DL79} in leading order: see figure~\ref{LARGE-T}. That is, they
are independent of $s$ and vary as $t^{-8}$.
There are many unanswered questions about this\cite{DL96}:
why does this simple behaviour set in already at such a small $t$, why
is it not significantly altered by higher-order perturbative corrections,
and are the data really energy-independent?
It will be interesting to check this at LHC energies.
A possibility
is that triple-gluon exchange will be replaced at higher energies with
triple-hard-pomeron exchange, so that the
large-$t$ differential cross section actually rises with increasing energy.

\begin{figure}[t]
\begin{center}
\epsfxsize=\hsize\epsfbox[25 580 310 770]{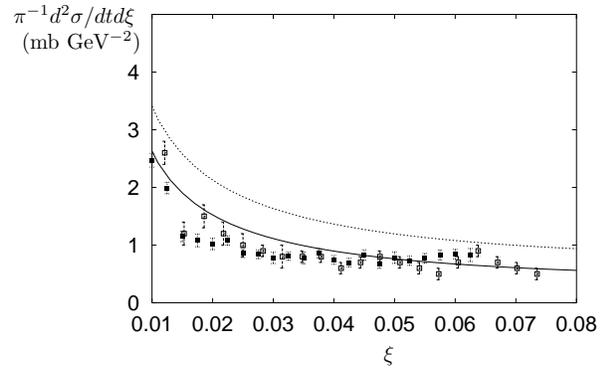}
\end{center}
\vskip -5truemm
\caption{Diffraction dissociation data at $\sqrt s=23$ GeV
(open points\ct{CHLM76})
and 630 GeV (black points\ct{Boz84a}) at $t=-0.75$ GeV$^2$.
The lower curve is from a simple model
and is for 23 GeV; the upper curve is the prediction for 630 GeV.}
\label{SEVENTYFIVE}
\end{figure}

\section{Soft diffraction dissociation}

We say that a $pp$ scattering event is diffractive if one of the
protons loses only an extremely small fraction $\xi$ of its momentum.
In diffraction dissociation, the other proton breaks up. The
mechanism by which this occurs is supposed to be pomeron exchange, as
is seen in figure~\ref{DIFFRACTION}. Although the pomeron is not a
particle, it is as if it collides with the second proton, and one talks
of the pomeron-proton cross section. This cross section should be similar 
to hadron-hadron-scattering cross sections. In particular, it should rise
with energy. But the $pp$ diffraction-dissociation data show no sign
of this rise. Figure~\ref{SEVENTYFIVE} shows data at $\sqrt s=23$
and 630 GeV and the curves are what is expected to result from the
rising pomeron-proton cross section. There is no agreed explanation
for this discrepancy, though there have been suggestions that, for some
reason, the pomeron flux does not show the expected behaviour with
increasing energy\cite{Gou95,ES00}.

\begin{figure}[t]
\vskip -4truemm
\begin{center}
\epsfxsize=0.4\hsize\epsfbox{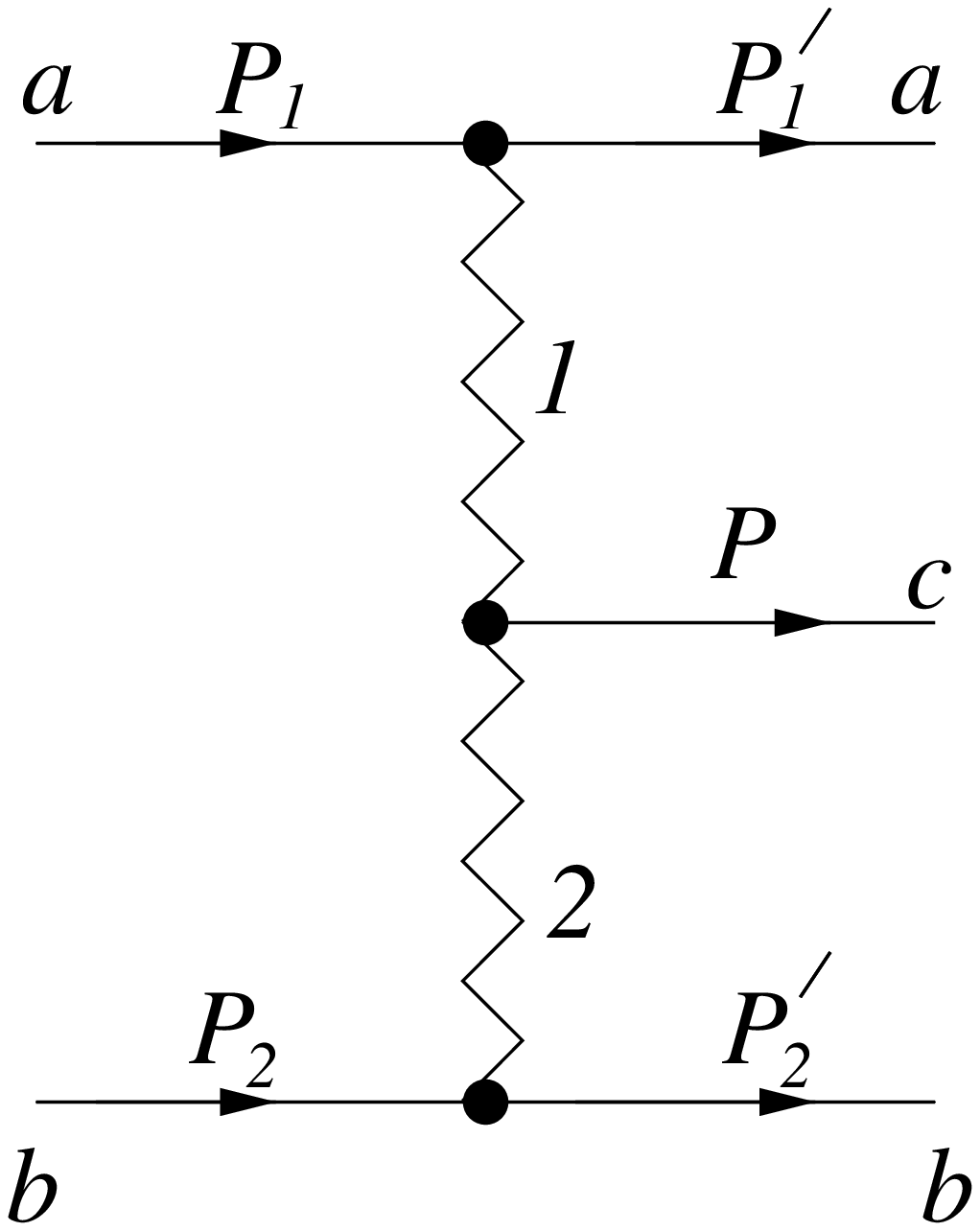}
\end{center}
\vskip -9truemm
\caption{Exclusive central production of a Higgs}
\label{PRODUCTION}o\vskip 9truemm
\vskip -6truemm
\begin{center}
\epsfxsize=0.4\hsize\epsfbox{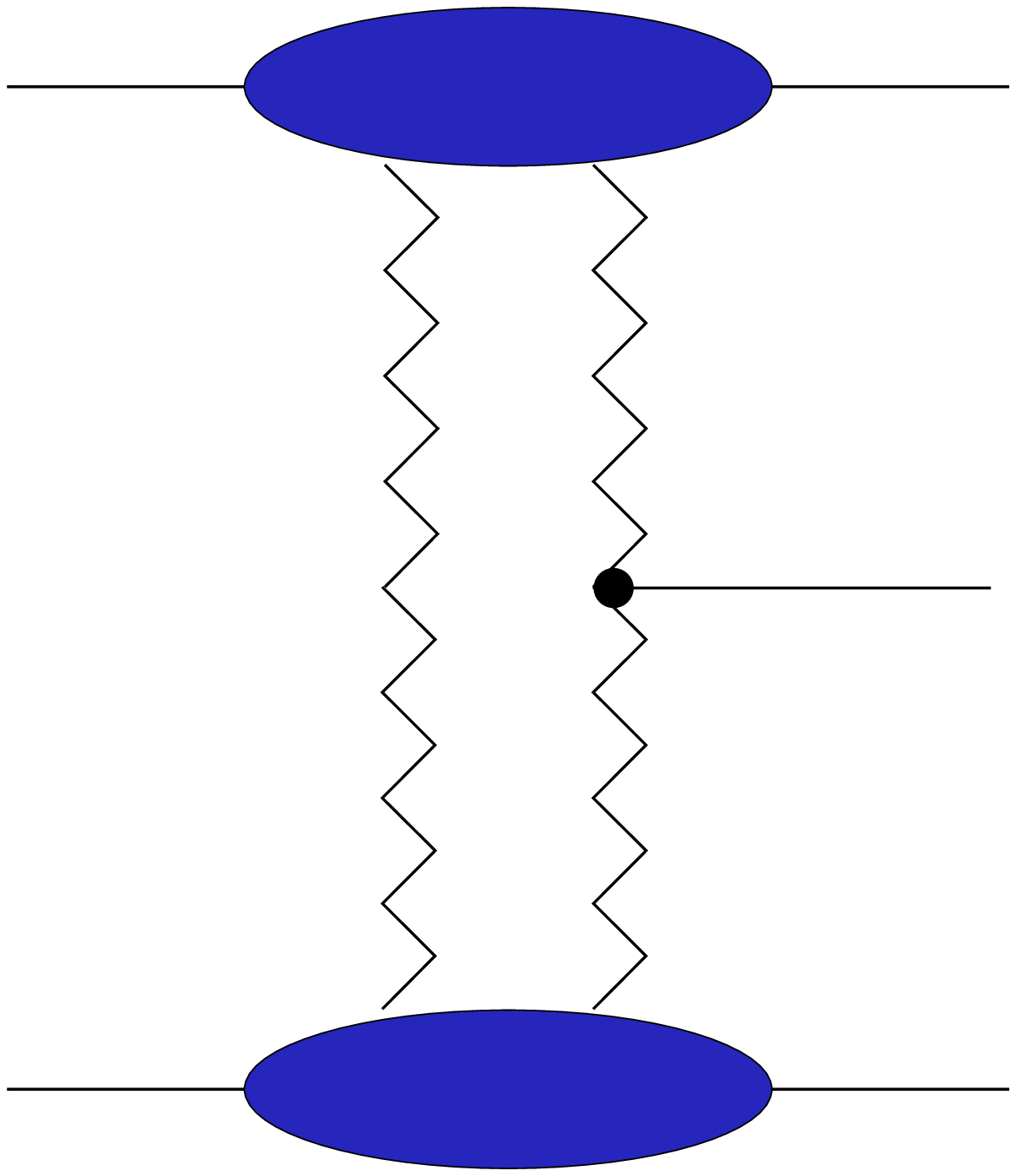}
\end{center}
\vskip -9truemm
\caption{Screening correction to figure~\ref{PRODUCTION}}
\label{HIGGSSCREEN}
\end{figure}

\section{Exclusive Higgs production}

The exclusive process $pp\to pHp$, where both the final-state protons
emerge with very high longitudinal momentum, has been discussed
extensively over the last ten years or so\cite{SNS90,BL91}. This
reaction should be generated by double pomeron exchange: see figure~%
\ref{PRODUCTION}.

Interest in the reaction has been revived by the suggestion\cite{AR00}
that it might be a good way to discover the Higgs. Higgs searches
in hadronic collisions have big background problems, but it is
argued that, by measuring  the momenta of the final-state protons
in figure~\ref{PRODUCTION} very accurately, one may determine the
missing mass very accurately. So one needs to integrate the background
only over a small mass range, so reducing its importance.

The argument now is whether the cross section for the process is large enough
to make it visible. In particular, are screening corrections so
large as to make the cross section very small? See figure~%
\ref{HIGGSSCREEN}. It has been claimed that indeed this is so and
that there is a suppression of more than an order of
magnitude. However, this claim is based not on a calculation of the
screening itself, but on an
argument that there is a very large likelihood that the two rapidity gaps
in the mechanism of figure~\ref{PRODUCTION} will be filled in
by the production of extra particles. But if one wants to 
calculate the amplitude for
a given process, it is not relevant what else might happen. If one
applied the same argument to pp elastic scattering one would conclude that
the cross section should be extremely small, when in fact it is more
than a quarter of the total cross section. It is true that in the eikonal
model screening corrections are related to the probability of filling
in the rapidity gap\cite{GLM}, but this is special to the eikonal model
and there are good reasons not to trust the eikonal model.
My own belief is that screening corrections as in figure~\ref{HIGGSSCREEN}
give a 50\% suppression at most. The argument is related to that over
whether Froissart-bound considerations have an important effect on
how large cross sections are allowed to be.  So I think that the cross section
for exclusive Higgs production is an order of magnitude bigger than
has been claimed recently\cite{KMR00}.

\begin{figure}[t]
\begin{center}
\epsfxsize=0.633\hsize\epsfbox[0 0 400 400]{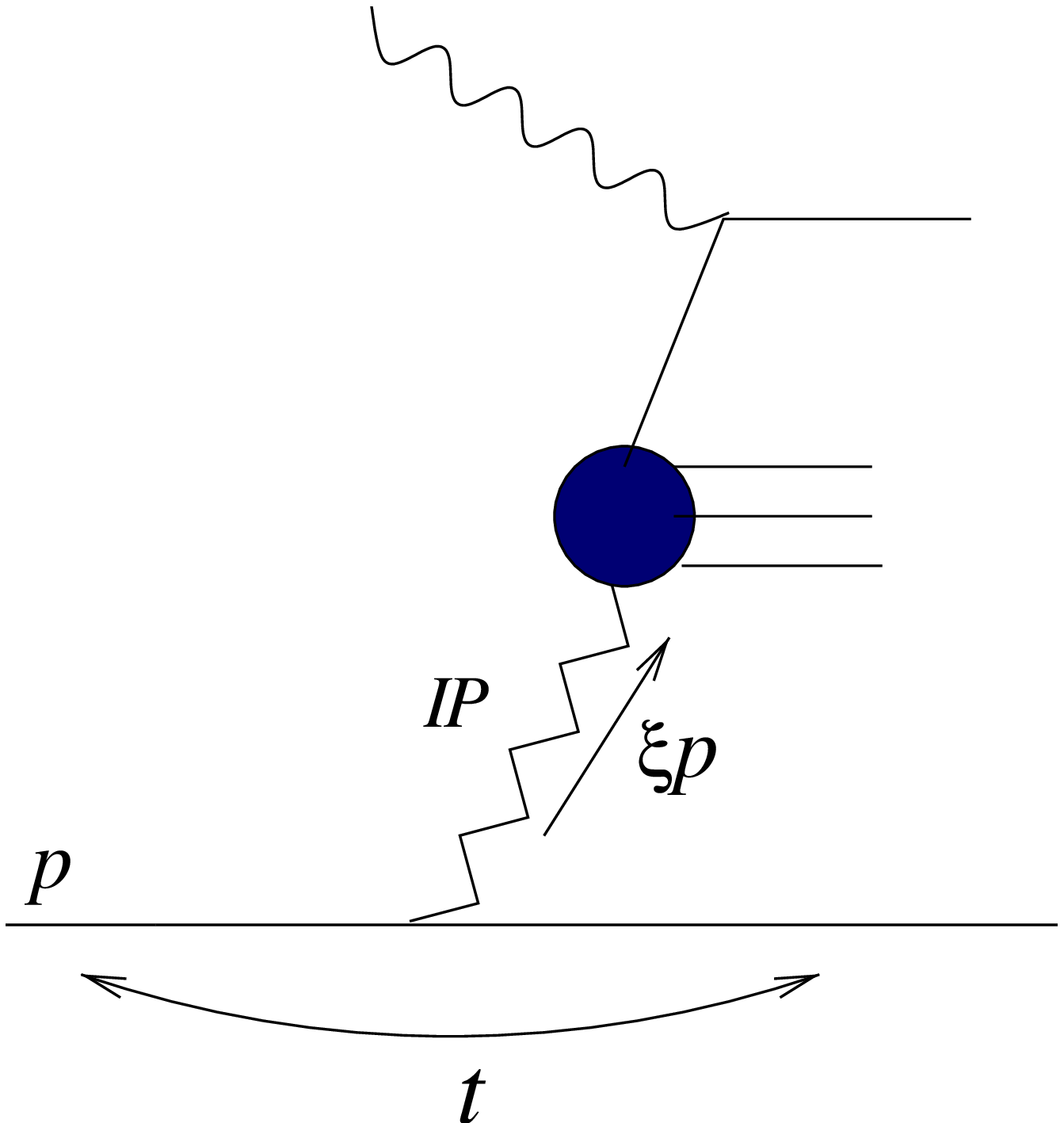}
\end{center}
\vskip -9truemm
\caption{Diffractive electroproduction}
\label{POMSTRUCT}
\vskip 9truemm
\begin{center}
\vskip 7truemm
\epsfxsize=\hsize\epsfbox[90 600 360 770]{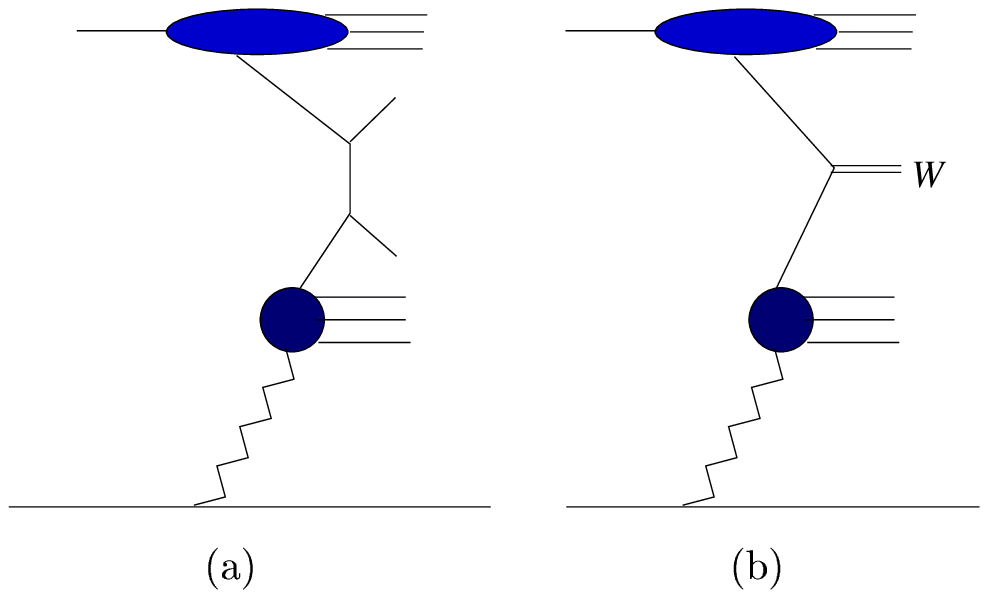}
\end{center}
\vskip -9truemm
\caption{$pp$ collisions with a very fast proton in the final state:
production of (a) a high-$P_T$ jet pair and (b) W boson.}
\label{POMSTRUCT2}
\end{figure}
\section{Hard diffraction}

The prediction\cite{IS85} that there should be a sizeable probability
that hard reactions also could lead to a very fast final-state
proton was first confirmed in an experiment\cite{UA888} at the CERN
$\bar pp$ collider. In $\gamma^*p$ scattering the mechanism is that shown
in figure~\ref{POMSTRUCT}. Although the pomeron is not a particle, it
is as if the mechanism involves a hard $\gamma^*$-pomeron collision and
so measures the structure function of the pomeron, just as $\gamma^*$-proton
collisions measure the structure function of the proton.
This has been studied extensively at HERA\cite{ZH102}, where at small $x$ some
10\% of the events are found to be diffractive.

The Tevatron experiments have measured the diffractive production of
dijets and of the $W$. The mechanism of figure~\ref{POMSTRUCT2} suggests
that the same pomeron structure function should be involved as in
diffractive electroproduction, and that therefore again some
10\% of dijet or $W$ events should be diffractive. The result
that is found is an order of magnitude smaller\cite{CDFD002}. Again this
has been blamed\cite{KMR00} on the filling in of the rapidity
gap by the production of additional particles. Unlike the exclusive Higgs
production I have discussed before, these are inclusive processes, for
which we have a much less well-defined theoretical formalism, so I
do not find this explanation implausible. Nevertheless, I
wonder whether things will be different at the much higher energy of the LHC.

\begin{figure}[p]
\bc
\vskip -9truemm
\epsfxsize=0.9\hsize\epsfbox[50 50 400 300]{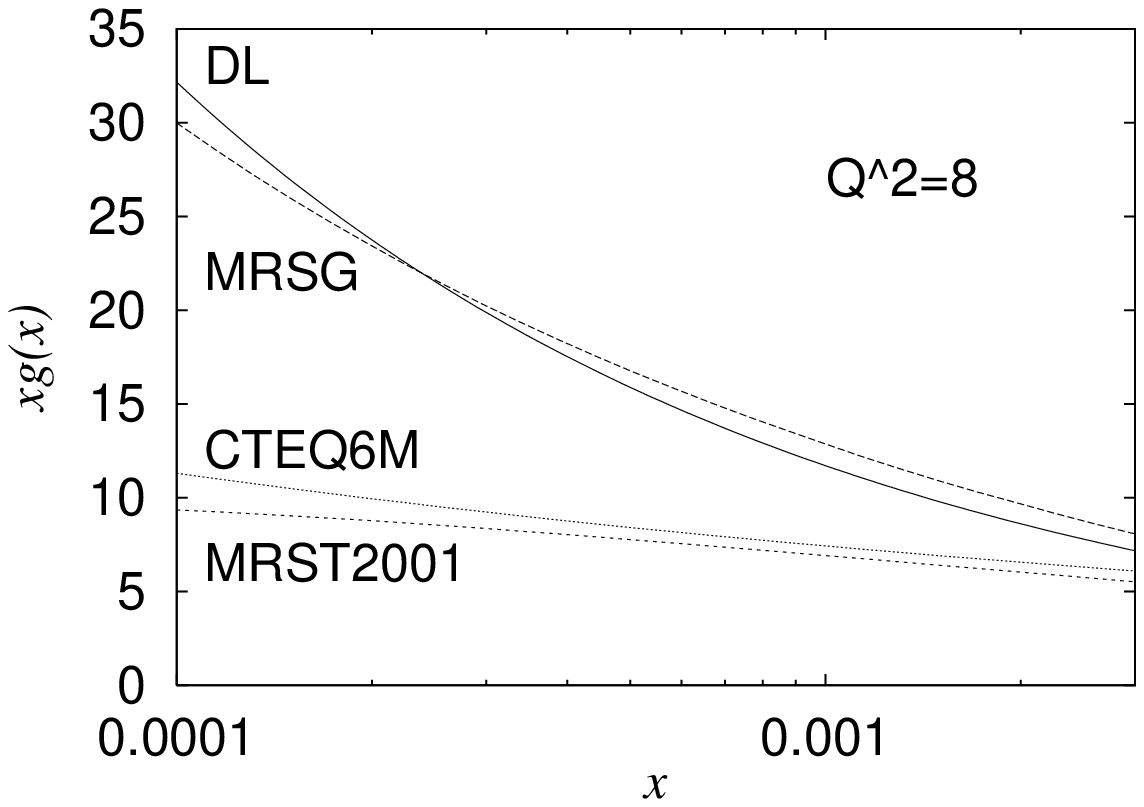}
\ec
\vskip -15truemm
\caption{Gluon structure functions\cite{dham,DL02}}
\label{GLUONSF}
\vskip 8truemm
\begin{center}
\epsfxsize=\hsize\epsfbox[100 300 410 760]{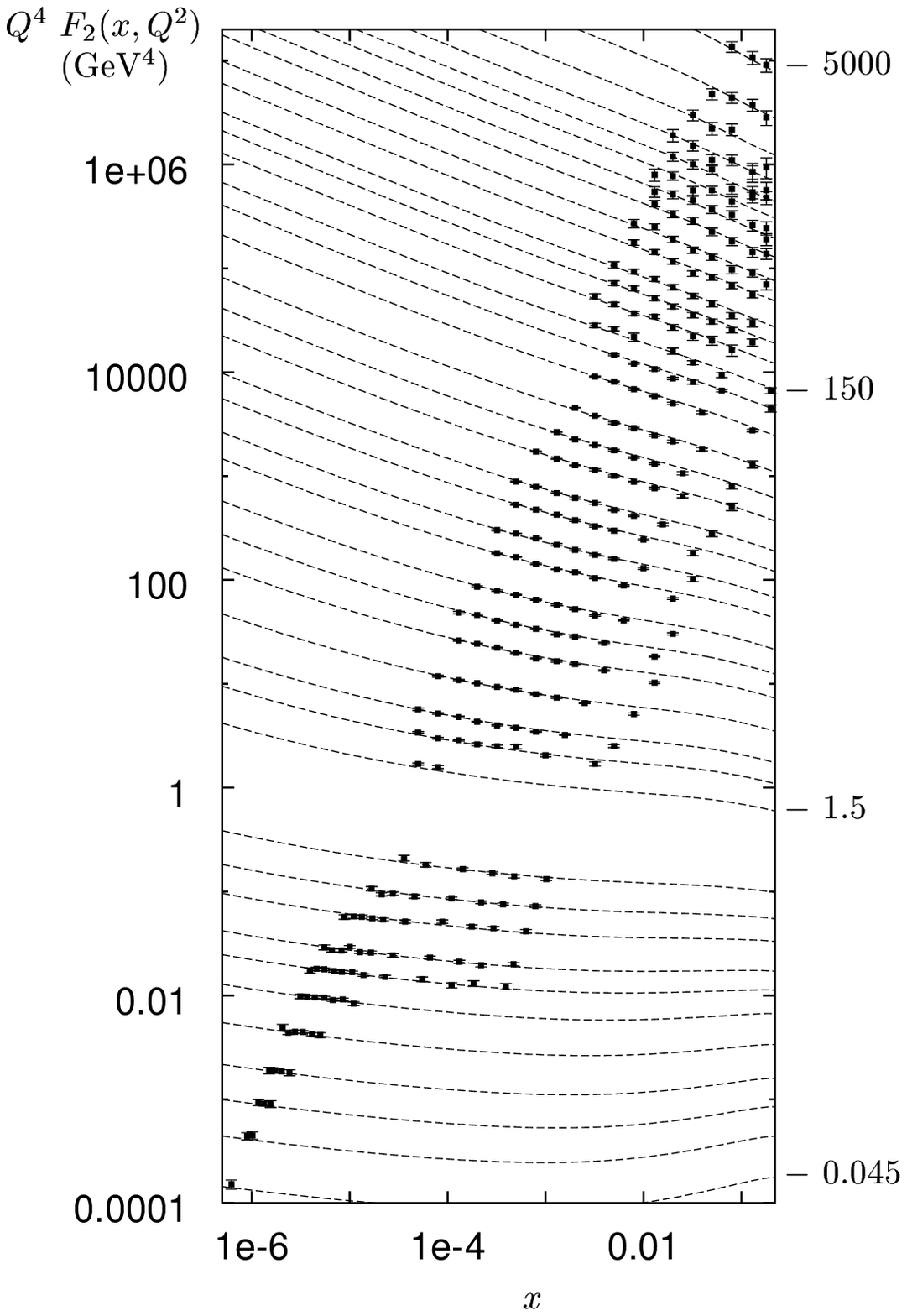}
\end{center}
\vskip -12truemm
\caption{Regge fit to ZEUS and H1 data for $F_2(x,Q^2)$ for $Q^2$ between 0.045
and 5000 GeV$^2$. The parameters were fixed using only data for
$x<0.001$ and therefore $Q^2\leq 35$ GeV$^2$.}
\label{F2}
\end{figure}

\section{Deep inelastic lepton scattering}

I have explained that do not understand how to apply DGLAP evolution 
at small $x$.
However, if we combine it with Regge theory and use an important message
from the HERA data for the charm structure function $F_2^c(x,Q^2)$, it
is possible\cite{DL02} reliably to extract the gluon structure function $g(x,Q^2)$
at small $x$. It turns out to be larger than nowadays is commonly believed.
This is seen in figure~\ref{GLUONSF}. The most recent CTEQ and MRST structure 
functions\cite{mrst,cteq}
agree well with each other and with those extracted by the two HERA
experiments\cite{H101,zeusfit} because they all use similar procedures; 
however, Donnachie
and I believe that the old MRSG structure function is nearer the truth.

When one tries to fit data, it is usually sensible to start with
the simplest assumptions and then refine them later. In its simplest
form, Regge theory leads to fixed powers of $x$ at small $x$, and it turns
out that two terms are enough: 
\be
F_2(x,Q^2)\sim f_0(Q^2)x^{-\epsilon_0}+f_1(Q^2)x^{-\epsilon_1}
\label{regge}
\ee
The second term corresponds to soft-pomeron exchange, with $\epsilon_1\approx
0.08$ determined from soft reactions. The data need a term that rises
more rapidly at small $x$; one needs $\epsilon_0\approx 0.4$. By fitting
the data at each $Q^2$, Donnachie and I found\cite{twopom} that
a successful and economical parametrisation of the coefficent functions
is provided by
$$
f_0(Q^2)=A_0 (Q^2)^{1+\epsilon_0}/(1+Q^2/Q_0^2)^{1+\epsilon_0/2}~~~~~~~~~~
$$
\begin{equation}
f_1(Q^2)=A_1 (Q^2)^{1+\epsilon_1}/(1+Q^2/Q_1^2)^{1+\epsilon_1}
\label{pheno}
\ee
with $Q_0\approx 3\hbox{ GeV}$ and $Q_1\approx 0.8\hbox{ GeV}$.
To make the fit, we used real-photon data and DIS data with $x\le 0.001$,
so that $Q^2$ ranges from 0.045 to 35 GeV$^2$. If we then simply multiply
the resulting form (\ref{regge}) by $(1-x)^7$, as is suggested by
the dimensional counting rules\cite{BF73b,MMT73}, it agrees quite well
with the HERA data even beyond $x=0.1$ and up to $Q^2=5000$ GeV$^2$.
This is shown in figure~\ref{F2}. Note that this factor $(1-x)^7$ should not
be taken too seriously; it is much too simple. 
\begin{figure}
\bc
\epsfxsize=\hsize\epsfbox[75 380 370 770]{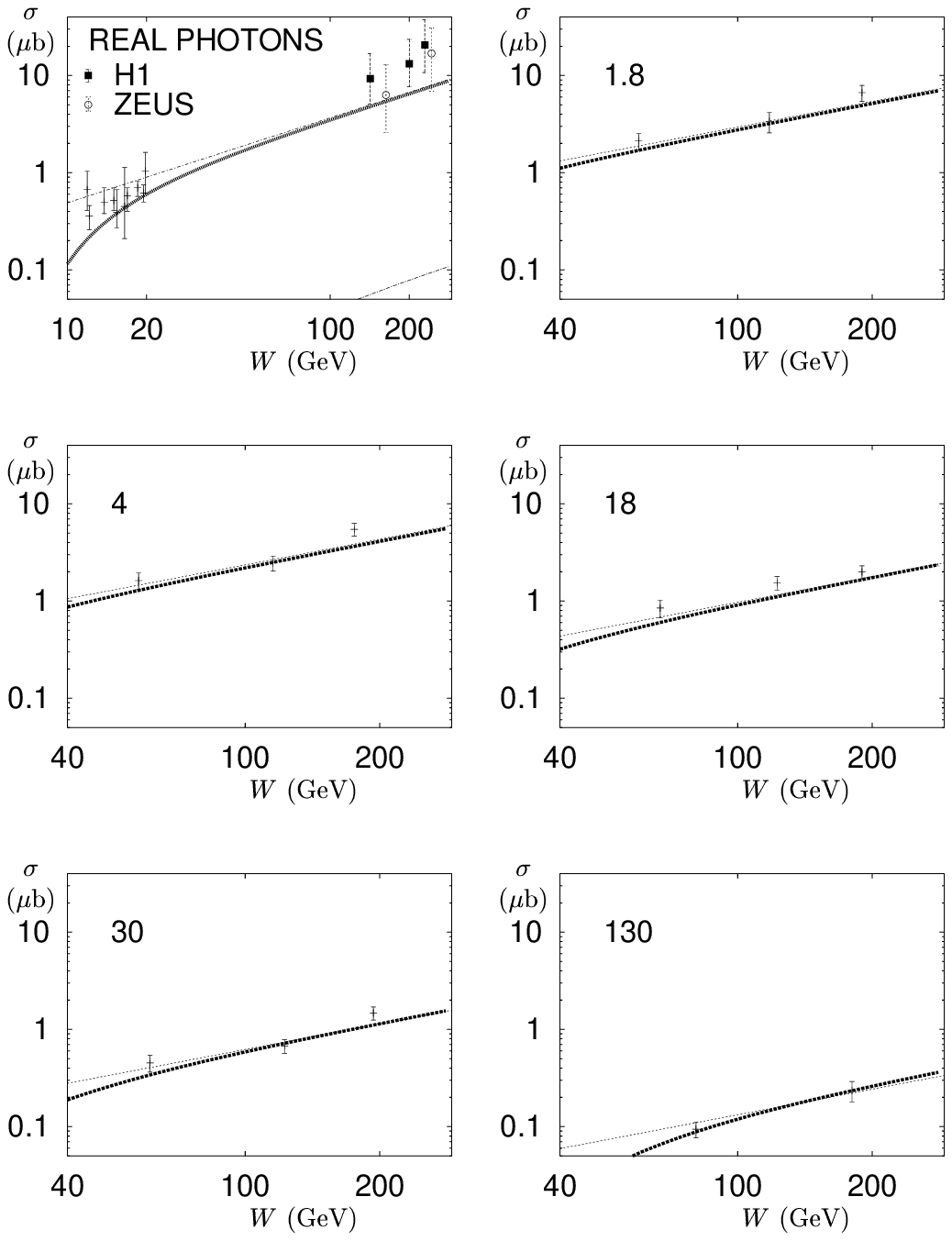}
\ec
\vskip -12truemm
\caption{Data\cite{zeusc} for the electroproduction of charm at various $Q^2$, with $W^{0.87}$ and pQCD fits (upper and lower curves, respectively)}
\label{CHARM}
\vskip 7truemm
\bc
\epsfxsize=0.69\hsize\epsfbox[75 560 350 765]{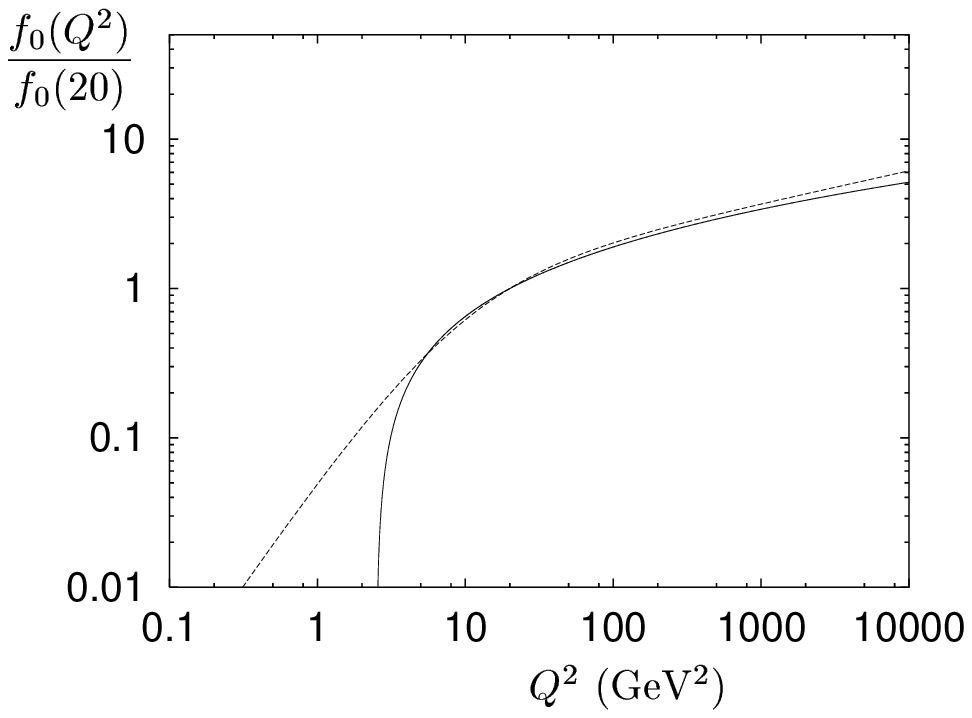}
\ec
\vskip -12truemm
\caption{pQCD evolution of the hard-pomeron coefficient function (solid curve)
with the phenomenological fit (broken curve)}
\label{evol}
\end{figure}

\def\one{{\textstyle {1\over 9}}}
\def\four{{\textstyle {4\over 9}}}
Data\cite{zeusc} for the charm structure function $F_2^c(x,Q^2)$ have the
remarkable property that, at all available $Q^2$, they fit to just
the single hard-pomeron power of $x$. Further, to an excellent approximation
the coupling of the hard pomeron appears to be flavour blind:
\be
F_2^c(x,Q^2)=f_c(Q^2)\,x^{-\epsilon_0}
\ee
with
$$
f_c(Q^2)=\four /(\four +\one +\one +\four )f_0(Q^2)= {0.4}~f_0(Q^2)
$$
\be
\ee
So if we define a charm-production cross section
\be
\sigma^c(W)={4\pi^2\alpha_{\hbox{{\fiverm EM}}}\over Q^2}F_2^c(x,Q^2)
\Big|_{x=Q^2/(W^2+Q^2)}
\ee
it behaves as $W^{2\epsilon_0}$ at all $Q^2$, even down to \hbox{$Q^2=0$}: see
figure~\ref{CHARM}.
Perturbative QCD directly relates $F_2^c(x,Q^2)$
to the gluon structure function, so that at small $x$ it too must be 
dominated by hard-pomeron exchange alone, even at quite small values of
$Q^2$. This is what causes the rapid rise  at small $x$ of the
DL curve in figure~\ref{GLUONSF}.

\def\u{{\bf u}}\def\P{{\bf P}}\def\f{{\bf f}}
\section{DGLAP evolution}

I have already explained that the usual procedure introduces
spurious singularities into the splitting matrix $\P$ that appears in
the DGLAP equation (\ref{dglap}).
My own belief is that $\P (N,\alpha_s(Q^2))$ has no singularities in
the complex-$N$ plane, or at least no relevant singularities. My reason
is that solving (\ref{dglap}) would cause a singularity of
$\P (N,\alpha_s(Q^2))$ to induce an {\it essential} singularity 
in $\u(N,Q^2)$ (that is, a nasty one). The variable $N$ is closely
related to the orbital angular momentum $\ell$, and I was brought up\cite{elop} 
to believe that matrix elements such as $\u(N,Q^2)$ do not have essential 
singularities in the complex $\ell$-plane. This point of view contrasts
with  that of those who believe that the value of $\epsilon_0$ is associated
with a singularity of $\P (N,\alpha_s(Q^2))$ and may even be calculated,
perhaps by refining the BFKL approach. I think that very probably
$\epsilon_0$ is a nonperturbative quantity that at present cannot be
calculated.

A fixed-power behaviour $x^{-\epsilon_0}$ of $F_2(x,Q^2)$, such as
in (\ref{regge}), corresponds to an $N$-plane pole:
\be
\u(N,Q^2)\sim {\f(Q^2)\over N-\epsilon_0}~~~~~~~\f(Q^2)=\Big(\matrix{f_0(Q^2)\cr f_g(Q^2)\cr}\Big )
\ee
If we insert this into the DGLAP equation
(\ref{dglap}) and equate the residue of the pole on each side
of the equation, we find
\be
{{\pd\over\pd t}\f(Q^2)= \P (N=\epsilon_0,\alpha_s(Q^2))\, \f(Q^2)}
\label{evolution}
\ee
$\epsilon_0$ is far enough from 0 for the expansion 
of $\P (N=\epsilon_0,\alpha_s(Q^2))$ to be reasonably safe. So we may
easily use the DGLAP equation to calculate the evolution of the
hard-pomeron component of $F_2(x,Q^2)$. But this is not the case for
the soft-pomeron component, because $\epsilon_1\approx 0.08$ is too
close to 0.

\begin{figure}[t]
\bc
\epsfxsize=0.9\hsize\epsfbox[50 50 390 290]{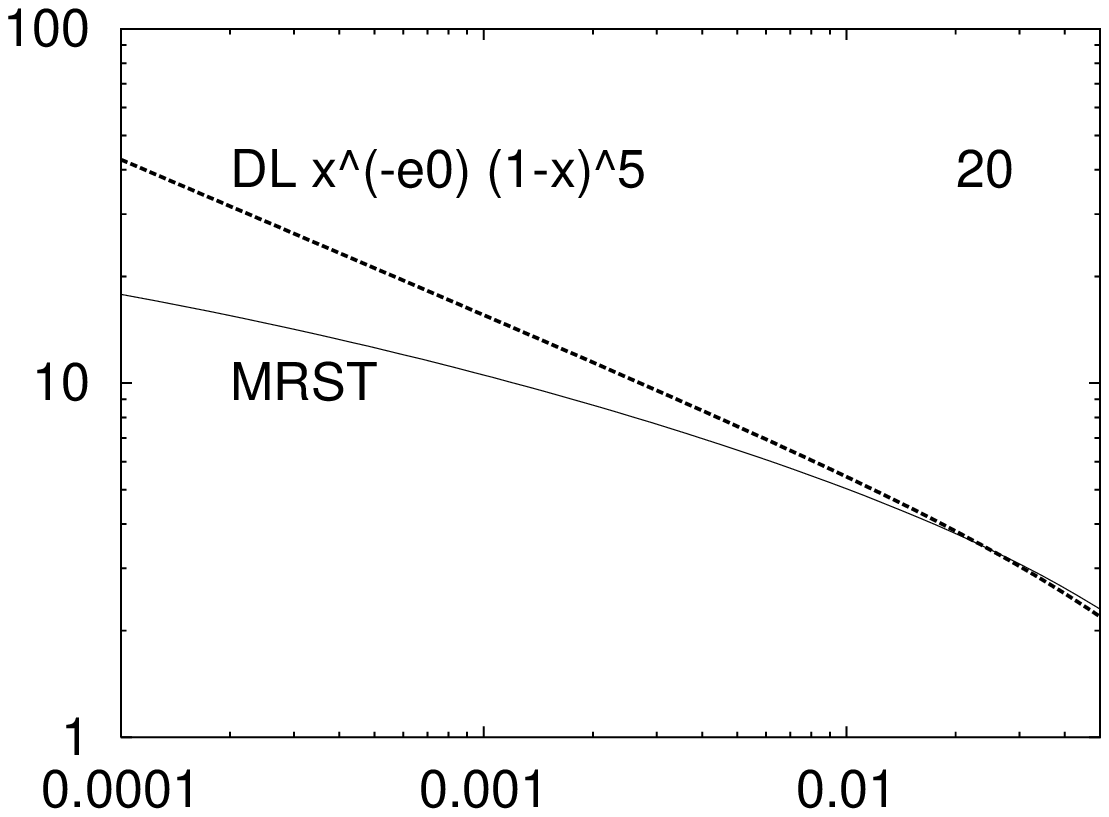}
\epsfxsize=0.9\hsize\epsfbox[50 50 390 290]{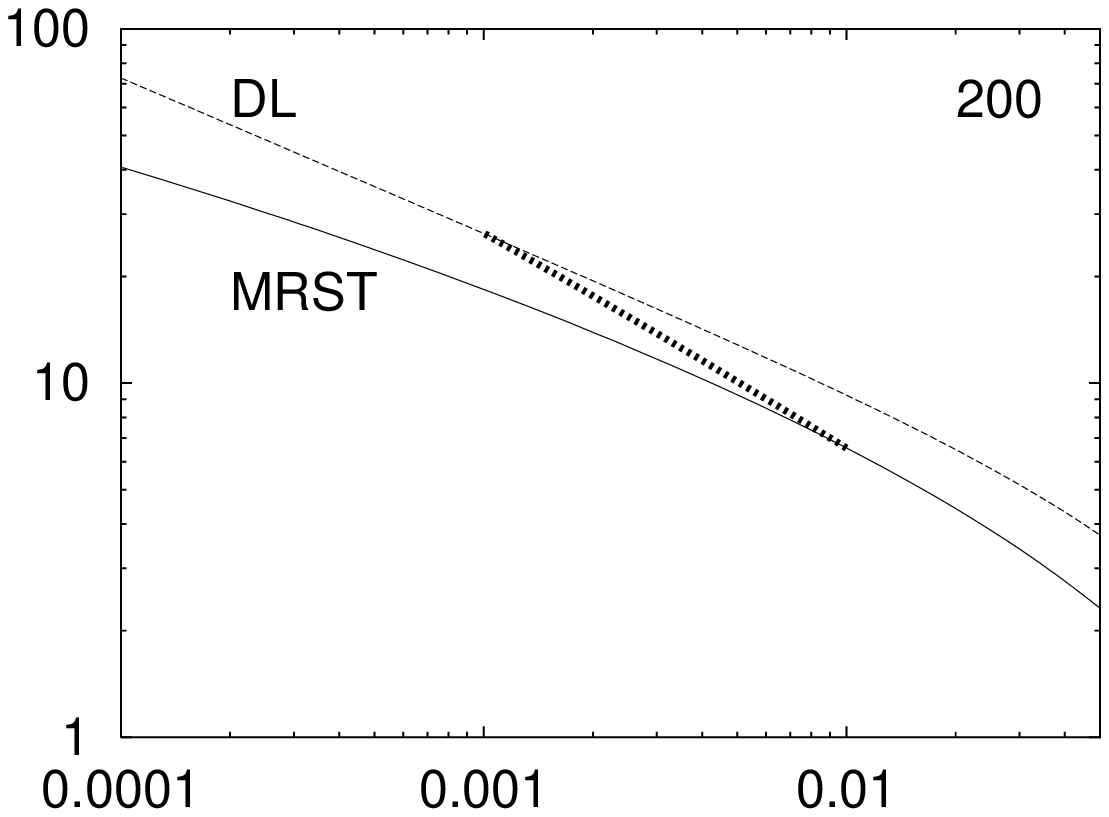}
\ec
\vskip -11truemm
\caption{Gluon structure function at $Q^2=20$ and 200 GeV$^2$}
\label{dlmrst}
\end{figure}

According to figure~\ref{GLUONSF}, the various gluon structure
functions come together at $x\approx 0.01$. It is reasonable to assume that
for values of $x$ larger than this the evolution of the two elements
of $\u(x,Q^2)$ does not use values of $N$ close to 0 and therefore the
conventional analysis is correct. 
So we can start at some not-too-large value of $Q^2$, 20 GeV$^2$ say.
We determine the value of $f_0(Q^2)$ there from the phenomenological fit
(\ref{pheno}) and $f_g(Q^2)$ from the MRST gluon structure function $xg(x,Q^2)$,
which for $x$ greater than about 0.01 fits very well to $x^{-\epsilon_0}
(1-x)^5$. 
We  choose $\Lambda$ such that
$\alpha_s(M_X^2)=0.116$ and use (\ref{evolution}) to calculate\cite{DL02} the 
evolution of 
$f_0(Q^2)$ and $f_g(Q^2)$ in both directions. The result for $f_0(Q^2)$ is the 
continuous curve in figure~\ref{evol}. The dashed curve is the 
phenomenological form (\ref{pheno}). Provided we adjust $\Lambda$
so that still $\alpha_s(M_X^2)=0.116$, LO and NLO evolution give almost 
identical results.

The agreement between the pQCD calculation and the phenomenological
curve is a success not only for the concept of the hard pomeron, but
also for pQCD itself. The evolution is from a single value of $Q^2$,
not the customary global fit\cite{mrst,cteq}, and it introduces far fewer
parameters.

Notice that, as $Q^2$ increases, the large-$x$ behaviour of $xg(x,Q^2)$
becomes steadily steeper than \hbox{$(1-x)^5$}, and so the largest value of $x$
for which $x^{-\epsilon_0}$ is a good approximation to the structure
function steadily decreases. Figure~\ref{dlmrst} shows an estimate
of this.

We may use the gluon structure function to calculate the charm structure
function $F_2^c(x,Q^2)$. The result, using just LO photon-gluon fusion
with a charm-quark mass $m_c=1.3$ GeV,
is the solid curves in figure~\ref{CHARM}. This is an important check
on the consistency of the approach. As is seen in figure~\ref{HARDCHARM},
a steep gluon distribution is needed to fit the data at small $Q^2$.

In conclusion,
the conventional approach to evolution needs modifying at small $x$.
It can be corrected if we combine it with Regge theory,
but only partly --- we can only treat the hard-pomeron part.
The resulting gluon distribution
is larger at small $x$ than has so far been supposed
and gives a good description of charm production.
I should add that we want good data for the longitudinal structure function,
because this gives the most direct window on the gluon distribution.

\def\medskip {\vskip 3truemm}
\section{Summary}

\b What physics explains the dramatic HERA\break effect?
\medskip
\b Is unitarity a constraint on hard collisions?
\medskip
\b Do $pp$ and/or $\gamma p$ total cross sections contain a hard term?
\medskip
\b Why do we see no odderon at $t=0$?
\medskip
\b How do we understand soft diffraction dissociation?
\medskip
\b Is diffractive Higgs production large enough to measure?
\medskip
\b Why does HERA see more hard diffractive events than the Tevatron?
\medskip
\b The conventional approach to evolution needs modifying at small $x$
\medskip
\b It can be corrected if we combine it with Regge theory
\medskip
\b But how do we handle the soft-pomeron part?
\medskip
\b The gluon distribution is larger at small $x$ than has so far been supposed
\medskip
\b \BrickRed{We want good data for the longitudinal structure function}

\def\bf{}

\end{document}